\documentclass[12pt]{article}
\usepackage{graphics}
\usepackage{epsfig}
\newcommand{\stttx}{$\tilde{g} \to  t\bar{t}\tilde{\chi}^0_1~$}
\newcommand{\stttxg}{$\tilde{g} \to  t\bar{t}\tilde{\chi}^0_1g~$}

\begin{document}
\title{  QCD correction to gluino decay to $t\bar{t}\tilde{\chi}^0_1$\\
 in the MSSM } \vspace{3mm}
\author{{ Duan Peng-Fei, Zhang Ren-You, Ma Wen-Gan, Guo Lei, and Zhang Yu }\\
{\small Department of Modern Physics, University of Science and Technology}\\
{\small of China (USTC), Hefei, Anhui, 230026, P.R.China}}
\date{}
\maketitle

\vskip 12mm
\begin{abstract}
\par
We calculate the complete next-to-leading order QCD corrections to
the three-body decay of gluino into top-pair associated with a
lightest neutralino in the minimal supersymmetric standard model. We
obtain that the LO and NLO QCD corrected decay widths of $\tilde{g}
\to t \bar{t} \tilde{\chi}_1^0$ at the ${\rm SPS~6}$ benchmark point
are $0.1490~GeV$ and $0.1069~GeV$ respectively, and the relative
correction is $-28.2\%$. We investigate the dependence of the QCD
correction to $\tilde{g} \to t \bar{t} \tilde{\chi}_1^0$ on
$\tan\beta$ and the masses of gluino, scalar top quarks and the
lightest neutralino around the ${\rm SPS6}$ benchmark point,
separately. We find that the NLO QCD corrections suppress the LO
decay width, and the absolute relative correction can exceed $30\%$
in some parameter space. Therefore, the QCD corrections to the
three-body decay $\tilde{g} \to t\bar{t}\tilde{\chi}^0_1$ should be
taken into account for the precise experimental measurement at
future colliders. Moreover, we study the distributions of the
top-pair invariant mass ($M_{t\bar t}$) and the missing energy
($E^{miss}$), and find that the line shapes of the LO distributions
of $M_{t\bar t}$ and $E^{miss}$ are not obviously distorted by the
NLO QCD corrections.
\end{abstract}


\vfill \eject \baselineskip=0.32in

\renewcommand{\theequation}{\arabic{section}.\arabic{equation}}
\renewcommand{\thesection}{\Roman{section}}
\newcommand{\nb}{\nonumber}
\makeatletter      
\@addtoreset{equation}{section}
\makeatother       

\section{Introduction}
\par
The supersymmetric theories \cite{MSSM} are the most motivated
extensions of the standard model (SM). They predict that the SM
particles have their corresponding superpartners (sparticles), and
the CERN Large Hadron Collider (LHC) and the
International Linear Collider (ILC) might provide the experimental
facilities to confirm the existence of these new particles. Among
all the supersymmetric models, the minimal supersymmetric standard
model (MSSM) with $R$-parity conservation is the most interesting
one which is well studied in recent years \cite{MSSM}. In the
$R$-conserving MSSM, there exists a stable lightest neutral
supersymmetric particle (LSP) which is called the lightest
neutralino denoted as $\tilde{\chi}_1^0$.

\par
Supersymmetry (SUSY) must be broken in the practical world and the
sparticle mass spectrum depends on the SUSY breaking mechanism.
The fundamental MSSM parameters need to be determined from
the precise measurement of the masses, production cross sections
and decay widths of these superpartners. With these parameters we
can reconstruct the SUSY breaking mechanism and probe the MSSM.

\par
Among all the sparticles in the MSSM, the two colored scalar quark
(squark) chiral eigenstates $\tilde q_{L}$ and $\tilde q_{R}$ are
the corresponding superpartners of the chiral quarks appearing in
the SM. The physical mass eigenstates $\tilde{q}_1$ and
$\tilde{q}_2$ are the mixtures of these two chiral eigenstates. The
scalar partners of top quarks are expected to be the lightest
squarks in supersymmetric theories. As the colored supersymmetric
particles, squarks  and gluino may be produced copiously in hadronic
collisions. After the (pair) production of these particles they are
expected to decay via cascade decay to lighter SUSY particles
associated with quarks and/or leptons. Gluino may have many decay
modes in the MSSM depending on the mass spectrum of SUSY particles
\cite{decay,model}. In general, gluino has three major decay
channels, i.e., (1) via an off-shell squark to two quarks and a LSP,
$\tilde{g} \to q \bar{q}+\tilde{\chi}_1^0$, (2) via an off-shell
squark to two quarks and a chargino or heavier neutralino, e.g.,
$\tilde{g} \to q \bar{q}'+\tilde{\chi}_1^+ $, and (3) through a
quark-squark loop to a gluon and a LSP, $\tilde{g} \to g+
\tilde{\chi}_1^0$. The lowest order decay width of the decay process
$\tilde{g} \to q\bar{q}\tilde{\chi}^0_1~$ was previously evaluated
in Refs.\cite{decay,tree-1,tree}. In Ref.\cite{tree-1} the authors
studied also the two-body decays $\tilde{g} \to g\tilde{\chi}^0_n$
including the one-loop QCD corrections by adopting the leading
logarithmic approximation. Recently, a complete study for all
two-body decay channels of the gluino in the complex MSSM with full
one-loop electroweak effects has been presented in
Ref.\cite{gluinodecay}. If $\tilde{t}_{1}$ is relatively lighter
than other squarks and the gluino is heavier ($m_{\tilde{g}} >
m_{\tilde{t}_1}$), the decay channel $\tilde{g} \to t
\bar{t}\tilde{\chi}_1^0 $ could have major contribution to the total
decay width of gluino \cite{tree-1,wanglt}. Therefore, the accurate
calculations including the NLO corrections to the three-body decay
\stttx are necessary.

\par
Currently, ATLAS and CMS experiments \cite{SUSY-ATLAS,SUSY-CMS} with
$4.7~fb^{-1}$ data have severely constrained the masses of the
strongly interacting SUSY particles --- squarks and gluinos.
However, in some SUSY scenarios these SUSY particles may still be
rather light. One interesting possibility is the light stop
scenario. In Ref.\cite{JBerger} the authors propose that the
recently developed techniques for tagging top jets can be used to
boost sensitivity of the LHC searches for the SUSY scenario in which
the third generation squarks are significantly lighter than those of
the first two generations.

\par
In this paper, we focus on the NLO QCD corrections to the three-body
decay \stttx. In section II we present the tree-level calculations
for the decay process \stttx. In section III we provide descriptions
of the analytical calculations of the NLO QCD corrections. In
section IV we present the numerical results around the scenario
${\rm SPS 6}$ point as proposed in the SPA project
\cite{SPA,snowmass}, and discuss the dependence of the QCD
correction on $\tan\beta$ and the masses of gluino, top squarks and
the lightest neutralino. Finally, we give a short summary.

\par
\section{LO calculations for \stttx}
\par
\begin{figure}[htbp]
\centering
\includegraphics[scale=1.0]{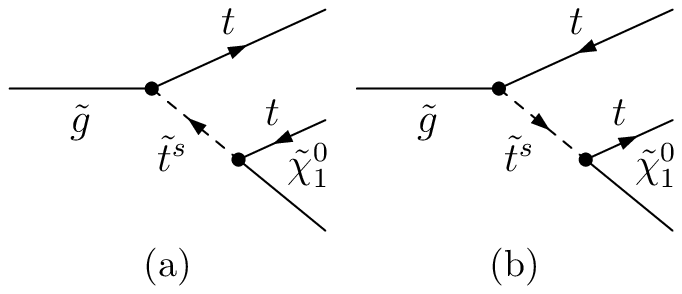}
\vspace*{0.3cm} \caption{ The tree-level Feynman diagrams for the
decay \stttx. } \label{Fig1}
\end{figure}

In this paper, we denote the decay process \stttx as
\begin{equation}
 \tilde{g}(p_1) \to t(p_2)+\bar{t}(p_3)+\tilde{\chi}^0_1(p_4),
\end{equation}
where $p_1$, $p_2$, $p_3$ and $p_4$ are the
four-momenta of the gluino and the decay products,
respectively. The leading order (LO) Feynman diagrams for this
decay are displayed in Fig.\ref{Fig1}, where $\tilde{t}_{s}$ with
the lower index $s$ running from 1 to 2 represent the two top
squarks $\tilde{t}_1$ and $\tilde{t}_2$.

\par
The LO decay width of the process \stttx can be obtained by using
the following formula:
\begin{equation}
\label{width} \Gamma_{LO} = \frac{(2\pi)^4}{2m_{\tilde{g}}} \int
\mathrm{d}\Phi_3 \overline{\sum}|{\cal M}_{tree}|^2.
\end{equation}
The summation is taken over the spins and colors of initial and
final states, and the bar over the summation recalls averaging over
the spin and color of initial gluino. $d\Phi_3$ is the
three-body phase space element defined as
\begin{eqnarray}
{d\Phi_3}=\delta^{(4)} \left( p_1-\sum_{i=2}^4 p_i \right)
\prod_{j=2}^4 \frac{d^3 \vec{p}_j} {(2 \pi)^3 2 E_j}.
\end{eqnarray}
After doing the integration in Eq.(\ref{width}) over the phase space,
we can get the expression of the tree-level decay width of the decay
process \stttx.

\par
In the LO calculations, the intermediate top squarks are potentially
resonant. We adopt the complex mass scheme (CMS) \cite{CMS} to deal
with the top squark resonance effect. In the CMS, the complex masses
of the unstable top squarks should be taken everywhere in both the
LO and NLO calculations. Then the gauge invariance is preserved and
the singularities of propagators for real $p^2$ are avoided. The
relevant complex masses are defined as
$\mu_{\tilde{t}_i}^2=m_{\tilde{t}_i}^2-im_{\tilde{t}_i}\Gamma_{\tilde{t}_i}$,
where $m_{\tilde{t}_i}~(i=1,2)$ are the conventional pole masses,
$\Gamma_{\tilde{t}_i}~(i=1,2)$ represent the corresponding total
widths of the top squarks, and the poles of propagators are located
at $\mu_{\tilde{t}_i}^2$ on the complex $p^2$-plane. Since the
unstable particles are involved in the loops for the ${\cal
O}(\alpha_s)$ QCD corrections, we shall meet the calculations of
$N$-point integrals with complex masses.

\vskip 5mm
\par
\section{NLO QCD corrections to \stttx }
\par
The calculations of the decay process \stttx in the MSSM are carried out in
t'Hooft-Feynman gauge. In the QCD NLO calculations, we use the dimensional
regularization (DR) scheme to isolate the ultraviolet (UV) and infrared (IR)
singularities. The Feynman diagrams and the relevant amplitudes are generated
by using FeynArts3.4 \cite{fey}, and the Feynman amplitudes are subsequently
reduced by FormCalc5.4 \cite{formloop}. The phase space integration is
implemented by using the Monte Carlo technique. The NLO QCD corrections to the
decay process \stttx, denoted as $\Delta\Gamma_{NLO}$, can be divided into two
parts: the virtual correction from one-loop diagrams ($\Delta\Gamma_{virtual}$) and
the real gluon emission correction ($\Delta\Gamma_{real}$), i.e.,
\begin{equation}
\Gamma_{NLO}=\Gamma_{LO}+\Delta\Gamma_{NLO}=\Gamma_{LO}+
\Delta\Gamma_{virtual}+\Delta\Gamma_{real}.
\end{equation}
Both the virtual and the real corrections contain IR singularities.
These  IR singularities exactly vanish after combining the virtual
correction with the real gluon emission correction together. Then
the decay width of \stttx including the NLO QCD corrections is
IR-finite.

\par
\subsection{Virtual corrections }
\vskip 5mm
{\bf A. Definitions of counterterms }
\par
\par
\begin{figure}[htb]
\centering
\includegraphics[scale=1.0]{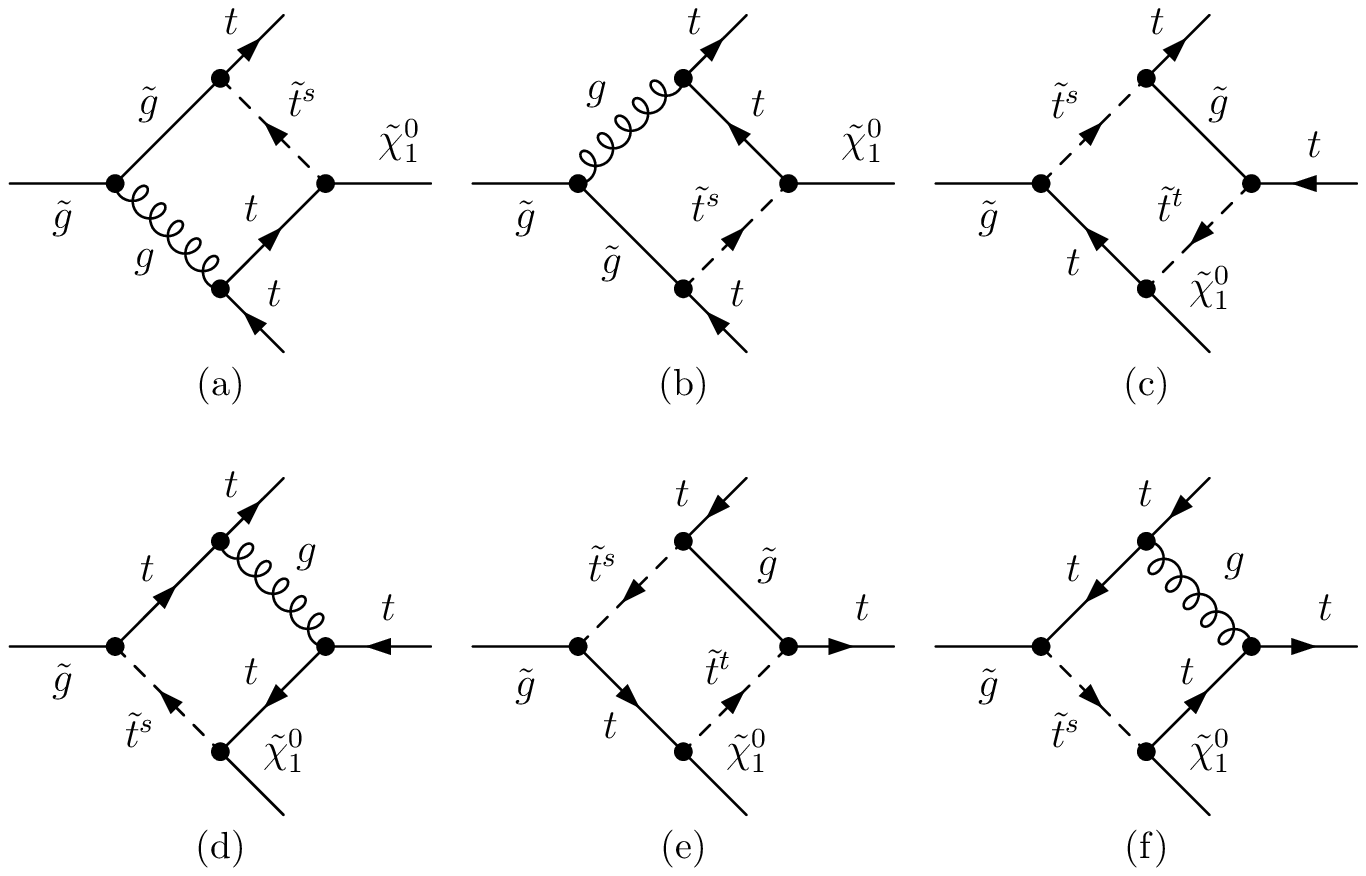}
\vspace*{0.3cm} \caption{ The QCD box diagrams for the decay process \stttx. } \label{Fig2}
\end{figure}

The renormalized virtual NLO QCD corrections to the decay \stttx
in the MSSM include the contributions from self-energy, vertex,
box, and the corresponding counterterm diagrams. We depict the QCD box
diagrams in Fig.\ref{Fig2} as a representative. The amplitude for the diagrams
with gluon in loops may contain both UV and soft IR singularities,
while the amplitude for the diagrams with gluino but
no gluon in loops contains only the UV singularities.
\par

\par
In order to remove the UV divergences, we employ the modified
minimal subtraction ($\overline{\rm MS}$) scheme \cite{Majhi} to
renormalize the strong coupling constant and the on-shell (OS)
scheme to renormalize the relevant colored fields and their masses,
respectively. After doing the renormalization procedure, the UV
singularities are eliminated. Here we list the definitions of the
counterterms of the wave functions of gluino, top squarks,
top-quark, the mixing matrix elements of top squark sector, the
strong coupling constant, the complex masses of top squarks and the
mass of top quark as
\begin{eqnarray} \label{def-1}
\tilde{g}_{L,0} &=& \left( 1 + \frac{1}{2} \delta Z_{L}^{\tilde{g}} \right) \tilde{g}_{L},
~~~\tilde{g}_{R,0} = \left( 1 + \frac{1}{2} \delta Z_{R}^{\tilde{g}} \right) \tilde{g}_{R}, \\
\label{def-2}\left(\begin{array}{c} \tilde{t}_{1,0} \\
\tilde{t}_{2,0}\end{array} \right)
&=&\left(\begin{array}{cc}1+\frac{1}{2}\delta Z^{\tilde{t}}_{11}& \frac{1}{2}\delta Z^{\tilde{t}}_{12} \\
\frac{1}{2}\delta Z^{\tilde{t}}_{21} & 1+\frac{1}{2}\delta
Z^{\tilde{t}}_{22}
\end{array} \right)\left(\begin{array}{c} \tilde{t}_1 \\
\tilde{t}_2 \end{array} \right),  \\
 \label{def-3}t_{L,0} &=& (1+\frac{1}{2}\delta Z^{t}_L)t_L, ~~~t_{R,0} = (1+\frac{1}{2}\delta
Z^{t}_R) t_R,  \\
\label{def-4} U_0^{\tilde t} &=& \left(\begin{array}{cc} U^{\tilde{t}}_{11,0}& U^{\tilde{t}}_{12,0}   \\
U^{\tilde{t}}_{21,0} & U^{\tilde{t}}_{22,0}
\end{array} \right) =
U^{\tilde t}+\delta U^{\tilde t}=\left(\begin{array}{cc} U^{\tilde{t}}_{11}& U^{\tilde{t}}_{12}   \\
\label{def-4} U^{\tilde{t}}_{21} & U^{\tilde{t}}_{22}
\end{array} \right) +
\left(\begin{array}{cc}\delta U^{\tilde{t}}_{11}& \delta U^{\tilde{t}}_{12} \\
\label{def-5} \delta U^{\tilde{t}}_{21} & \delta U^{\tilde{t}}_{22}
\end{array} \right),   \\
\label{def-6} \quad g_{s,0} &=& g_s+\delta g_s,~~~
m_{\tilde{t}_i,0}^2=\mu_{\tilde{t}_i}^2+\delta
\mu_{\tilde{t}_i}^2,~~(i=1,2 ),  \\
\nb m_{t,0} &=& m_t + \delta m_t,  \label{def-7}
\end{eqnarray}
where $U_0^{\tilde t}$ and $U^{\tilde t}$ are the bare and
renormalized $2 \times 2$ mixing matrices of the top squark sector,
$\tilde{t}_{i,0}$ and $\tilde{t}_i$ are the bare and renormalized
fields of top squarks, and $\delta Z^{\tilde{t}}_{ij}~(i,j=1,2)$ are the
renormalization constants of top quark fields.

\par
In above counterterm definitions, we split the real bare mass of top
squark squared ($m_{\tilde{t}_i,0}^2$) into complex renormalized
mass and mass counterterm  (i.e., $\mu_{\tilde{t}_i}^2$ and $\delta
\mu_{\tilde{t}_i}^2$). The bare fields of top squarks are split in
complex field renormalization constants and renormalized fields.
Similarly, $U^{\tilde{t}}_{ij,0}$ is separated into renormalization
constant part and renormalized matrix element which can be complex.
As a consequence, the renormalized Lagrangian, i.e. the Lagrangian
in terms of renormalized fields and parameters without counterterms,
is not hermitian, but the total Lagrangian keeps hermitian.

\vskip 5mm {\bf B. Complex renormalization for top squark sector }
\par
Because we use the CMS to deal with the possible top squark
resonance, the normal one-loop integrals must be continued onto the
complex plane. The formulas for calculating the IR-divergent
integrals with complex internal masses in the dimensional
regularization scheme are obtained by analytically continuing the
expressions in Ref.\cite{Stefan} onto the complex plane. The
numerical evaluations of IR-safe $N$-point ($N=1,2,3,4$) integrals
with complex masses, are implemented by using the expressions
analytically continued onto the complex plane from those presented
in Ref.\cite{OneTwoThree}. In this way, we created our in-house
subroutines to isolate analytically the IR singularities in
integrals and calculate numerically one-loop integrals with complex
masses based on the {\sc LoopTools-2.4} package \cite{formloop,ff}.
The analytic expression for scalar one-loop 4-point integral in the
complex mass scheme is also given in Ref.\cite{four}, we make a
careful comparison between ours and that provided in Ref.\cite{four}. It
shows that they are in good agreement.

\par
In our calculation the external gluino and top quark are real
particles, and their QCD on-shell self energies do not involve any
absorptive parts. But they become complex via the complex top squark
masses. The renormalization of top squark sector is more
complicated. It involves the renormalization of the mixing of the
two top squarks.

\par
With the renormalization conditions of the complex OS scheme
\cite{denner,Renor-1}, the renormalization constants of the complex
masses and wave functions of top squarks are expressed as
\begin{eqnarray}\label{eq2}
\delta\mu^2_{\tilde{t}_i}  =
\Sigma^{\tilde{t}}_{ii}(\mu_{\tilde{t}_i}^2) ,~~~~ \delta
Z_{ii}^{\tilde{t}} = -
\Sigma_{ii}^{\tilde{t}\prime}(\mu_{\tilde{t}_i}^2),~~~~\delta
Z_{ij}^{\tilde{t}} =
\frac{2\Sigma_{ij}^{\tilde{t}}(\mu_{\tilde{t}_j}^2)}
{\mu_{\tilde{t}_i}^2-\mu_{\tilde{t}_j}^2},~~(~i,j=1,2, i\neq j).
\end{eqnarray}

\par
We perform Taylor series expansions about real arguments for the top
squark self-energy as
\begin{eqnarray}\label{eq3}
\Sigma_{ij}^{\tilde{t}}(\mu_{\tilde{t}_j}^2)&=&\Sigma_{ij}^{\tilde{t}}
(m_{\tilde{t}_j}^2)+(\mu_{\tilde{t}_j}^2-m_{\tilde{t}_j}^2)
\Sigma_{ij}^{\tilde{t}\prime}(m_{\tilde{t}_j}^2)\;+\; {\cal O}\left(\alpha_s^3\right)~~\nb \\
&=&\Sigma_{ij}^{\tilde{t}}(m_{\tilde{t}_j}^2)-i
\textit{m}_{\tilde{t}_i}\Gamma_{\tilde{t}_j}
\Sigma_{ij}^{\tilde{t}\prime}(m_{\tilde{t}_j}^2)\;+\; {\cal
O}\left(\alpha_s^3\right),~~(i,j=1,2),
\end{eqnarray}
where $\Sigma^{\tilde{t}\prime}_{ij}(m_{\tilde{t}_j}^2)\equiv
\frac{\partial \Sigma^{\tilde{t}}_{ij}(p^2) }{\partial
p^2}|_{p^2=m_{\tilde{t}_j}^2}$, $\Sigma^{\tilde{t}}_{ij}\sim
{{\cal{O}}(\alpha_s)}$ and
$(\mu_{\tilde{t}_j}^2-m_{\tilde{t}_j}^2)\sim {{\cal{O}}(\alpha_s)}$.
By neglecting the higher order terms and using $\mu_{\tilde{t}_i}^2=
m_{\tilde{t}_i}^2-i m_{\tilde{t}_i}\Gamma_{\tilde{t}_i}$, we get
approximately the mass and wave function renormalization
counterterms of scalar top squarks as
\begin{eqnarray}
\label{eq4-1} \delta \mu^2_{\tilde{t}_i}  &=&
\Sigma_{ii}^{\tilde{t}}(m_{\tilde{t}_i}^2) +
(\mu^2_{\tilde{t}_i}-m_{\tilde{t}_i}^2)
\Sigma_{ii}^{\tilde{t}\prime}(m_{\tilde{t}_i}^2) ,~~ \\
\label{eq4-2} \delta Z_{ii}^{\tilde{t}} &=& -
\Sigma_{ii}^{\tilde{t}'}(m_{\tilde{t}_i}^2) ,   \\
\label{eq4-3}\delta Z_{ij}^{\tilde{t}} &=& \frac{2
\Sigma_{ij}^{\tilde{t}}(m_{\tilde{t}_j}^2)}
{m_{\tilde{t}_i}^2-m_{\tilde{t}_j}^2},~~(~i,j=1,2, i\neq j).
\end{eqnarray}

\par
By adopting the unitary condition for the bare and renormalized
mixing matrices of top squark sector, $U^{\tilde{t}}_0$ and
$U^{\tilde{t}}$, we obtain the expression for the counterterm of top
squark mixing matrix $\delta U^{\tilde{t}}$ as
\begin{eqnarray}\label{eq25}
\delta U^{\tilde t} &=&\frac{1}{4} \left(\delta Z^{\tilde{t}}-\delta
Z^{\tilde{t}\dagger} \right)U^{\tilde{t}}.
\end{eqnarray}
With the definitions shown in Eqs.(\ref{def-2}) and (\ref{def-4}),
the counterterms of the top squark mixing matrix elements can be
written in the following form:
\begin{eqnarray}\label{eq-26}
\delta U^{\tilde t}_{ij} &=& \frac{1}{4} \sum_{k=1}^2\left(\delta
Z^{\tilde{t}}_{ik} - \delta Z^{\tilde{t}\ast}_{ki}
\right)U^{\tilde{t}}_{kj} ~~(i,j=1,2).
\end{eqnarray}

\par
The explicit expressions for unrenormalizad self-energies of top
squarks, $\Sigma_{ij}^{\tilde{t}}(p^2)~(i,j=1,2)$, are written as
\begin{eqnarray}\label{self-1}
\Sigma_{ii}^{\tilde{t}}(p^2)&=&-\frac{\alpha_s}{3\pi}\left( 4
A_0[m_t^2]+A_0[\mu_{\tilde{t}_i}^2]\right)~~\nonumber \\
&+&\frac{4\alpha_s}{3\pi}\left[m_t
m_{\tilde{g}}(U_{i2}^{\tilde{t}}U_{i1}^{\tilde{t}\ast}
+U_{i1}^{\tilde{t}}U_{i2}^{\tilde{t}\ast})-m_{\tilde{g}}^2\right]
B_0[p^2,m_{\tilde{g}}^2,m_t^2]~~\nonumber \\
&-&\frac{4\alpha_s}{3\pi}p^2 \left(B_0[p^2,0,\mu^2_{\tilde{t}_i}]
+B_1[p^2,0,\mu^2_{\tilde{t}_i}]
+B_1[p^2,m_{\tilde{g}}^2,m_t^2] \right)~~\nonumber \\
&+&\frac{\alpha_s}{3\pi}\sum_{k=1,2}(U_{k1}^{\tilde{t}}U_{i1}^{\tilde{t}\ast}
-U_{k2}^{\tilde{t}}U_{i2}^{\tilde{t}\ast})(U_{i1}^{\tilde{t}}U_{k1}^{\tilde{t}\ast}
-U_{i2}^{\tilde{t}}U_{k2}^{\tilde{t}\ast})A_0[\mu_{\tilde{t}_k}^2],
\end{eqnarray}
\begin{eqnarray}\label{self-2}
\Sigma_{ij}^{\tilde{t}}(p^2)&=&\frac{4\alpha_s}{3\pi}m_t
m_{\tilde{g}}(U_{i2}^{\tilde{t}}U_{j1}^{\tilde{t}\ast}
+U_{i1}^{\tilde{t}}U_{j2}^{\tilde{t}\ast})B_0[p^2,m_{\tilde{g}}^2,m_t^2]~~\nonumber \\
&+&\frac{2\alpha_s}{3\pi}\sum_{k=1,2}(U_{k1}^{\tilde{t}}U_{j1}^{\tilde{t}\ast}
-U_{k2}^{\tilde{t}}U_{j2}^{\tilde{t}\ast})(U_{i1}^{\tilde{t}}U_{k1}^{\tilde{t}\ast}
-U_{i2}^{\tilde{t}}U_{k2}^{\tilde{t}\ast})A_0[\mu_{\tilde{t}_k}^2],\nb \\
&&~~~~~~~~~~~~~~~~~~~~~~(i\neq j,~i,j=1,2).
\end{eqnarray}

\par
From Eqs.(\ref{eq4-2}), (\ref{eq4-3}), (\ref{self-1}) and
(\ref{self-2}), we can obtain $Im(\delta Z_{ij}^{\ast})=Im(\delta
Z_{ji})$. The Eq.(\ref{eq-26}) tells us that we can choose both the
renormalized mixing matrix elements ($U^{\tilde{t}}_{ij}$) and their
counterterms ($\delta U^{\tilde{t}}_{ij}$) made up of real matrix
elements. Then the matrix $U^{\tilde{t}}$ can be expressed explicitly as
\begin{eqnarray}
U^{\tilde t} &=& \left(\begin{array}{cc} \cos\theta_t &\sin\theta_t   \\
-\sin\theta_t & \cos\theta_t
\end{array} \right).
\end{eqnarray}
.

\vskip 5mm
{\bf C. Renormalizations for $\tilde{g}$ and $\alpha_s$ }
\par
Since in our LO and NLO calculations we adopt the complex mass
scheme, the conventional top squark masses are replaced by the
renormalized top squark complex masses $\mu_{\tilde{t}_i}~(i=1,2)$
everywhere, including in the expressions for the gluino self-energy
and counterterm of the strong coupling constant. The renormalized
one-particle irreducible two-point Green function
of gluino is defined as follow
\begin{eqnarray}\label{7}
i\hat\Gamma^{\tilde{g}}(p)=i(\not{p}-m_{\tilde{g}}) +
i[\not{p}P_L\hat\Sigma^{\tilde{g}}_L(p^2)+\not{p}P_R\hat\Sigma^{\tilde{g}}_R(p^2)+P_L\hat\Sigma^{\tilde{g}}_{SL}(p^2)
+P_R\hat\Sigma^{\tilde{g}}_{SR}(p^2)],
\end{eqnarray}
where $P_{L,R}\equiv \frac{1}{2}(1\mp \gamma_5)$,
$\hat\Sigma^{\tilde{g}}_L$, $\hat\Sigma^{\tilde{g}}_R$,
$\hat\Sigma^{\tilde{g}}_{SL}$ and $\hat\Sigma^{\tilde{g}}_{SR}$ are
renormalized gluino self-energies. By taking $u$-, $d$-, $c$-, $s$-,
$b$-quarks being massless, we express the corresponding
unrenormalized gluino self-energies explicitly as,
\begin{eqnarray}\label{self-5}
\Sigma^{\tilde{g}}_L(p^2)&=&\Sigma^{\tilde{g}}_R(p^2)=
-\frac{3\alpha_s}{4\pi}+\frac{3\alpha_s}{2\pi}B_1[p^2,0,m^2_{\tilde{g}}]
-\frac{\alpha_s}{4\pi}\sum^{\tilde{q}=\tilde{u},
\tilde{d},\tilde{c},\tilde{s},\tilde{b}}_{i=1,2}
B_1[p^2,0,m^2_{\tilde{q}_i}]~~\nonumber \\
&-&\frac{\alpha_s}{4\pi}\sum_{i=1,2}B_1[p^2,m^2_{t},\mu^2_{\tilde{t}_i}],
\end{eqnarray}
\begin{eqnarray}\label{self-7}
\Sigma^{\tilde{g}}_{SL}(p^2) &=&\frac{3\alpha_s}{2\pi}m_{\tilde{g}}
-\frac{3\alpha_s}{\pi}m_{\tilde{g}}B_0[p^2,0,m^2_{\tilde{g}}]-
\frac{\alpha_s}{2\pi}\sum_{i=1,2}m_{t}U_{i1}^{\tilde{t}}U_{i2}^{\tilde{t}\ast}
B_0[p^2,m^2_{t},\mu^2_{\tilde{t}_i}],
\end{eqnarray}
\begin{eqnarray}\label{self-8}
\Sigma^{\tilde{g}}_{SR}(p^2) &=&\frac{3\alpha_s}{2\pi}m_{\tilde{g}}
-\frac{3\alpha_s}{\pi}m_{\tilde{g}}B_0[p^2,0,m^2_{\tilde{g}}]-
\frac{\alpha_s}{2\pi}\sum_{i=1,2}m_{t}U_{i2}^{\tilde{t}}U_{i1}^{\tilde{t}\ast}
B_0[p^2,m^2_{t},\mu^2_{\tilde{t}_i}].
\end{eqnarray}
By adopting the OS scheme to the external gluino, the wave function
renormalization constants of gluino can be fixed by using the
following formulas \cite{denner, hollik}:
\begin{eqnarray}
\delta
Z^{\tilde{g}}_L&=&-\left[\Sigma^{\tilde{g}}_L(m_{\tilde{g}}^2) +
m_{\tilde{g}}^2(\Sigma^{\tilde{g}'}_{L}(m_{\tilde{g}}^2)+
\Sigma^{\tilde{g}'}_{R}(m_{\tilde{g}}^2))
+ m_{\tilde{g}} (\Sigma^{\tilde{g}'}_{SL}(m_{\tilde{g}}^2) +
\Sigma^{\tilde{g}'}_{SR}(m_{\tilde{g}}^2))\right], \nb \\
\delta Z^{\tilde{g}}_R&=&-\left[\Sigma^{\tilde{g}}_R(m_{\tilde{g}}^2) +
m_{\tilde{g}}^2(\Sigma^{\tilde{g}'}_{L}(m_{\tilde{g}}^2)+
\Sigma^{\tilde{g}'}_{R}(m_{\tilde{g}}^2)) + m_{\tilde{g}}
(\Sigma^{\tilde{g}'}_{SL}(m_{\tilde{g}}^2) +
\Sigma^{\tilde{g}'}_{SR}(m_{\tilde{g}}^2))\right],
\end{eqnarray}
where $\Sigma^{\tilde{g}\prime}(m_{\tilde{g}}^2)\equiv
\frac{\partial \Sigma^{\tilde{g}}(p^2) }{\partial
p^2}|_{p^2=m_{\tilde{g}}^2}$.

\par
For the renormalization of the strong
coupling constant $g_s$, we adopt the $\overline{MS}$ scheme at the
renormalizatiion scale $\mu_r$, except that the divergences
associated with the top quark and colored SUSY particle loops are
subtracted at zero momentum \cite{gs}. We define the counterterm of
the strong coupling constant $\delta g_s$ as a summation of the
SM-like QCD term and SUSY-QCD term (i.e, $\delta g_s = \delta
g_s^{(SM-like)} + \delta g_s^{(SQCD)}$), and these two terms can be
expressed as
\begin{eqnarray}\label{gs-1}
\frac{\delta g_s^{(SM-like)}}{g_s} &=&
-\frac{\alpha_s(\mu_r)}{4\pi}\bigg{[}\frac{\beta_0^{(SM-like)}}
{2}\frac{1}{\bar{\epsilon}}+\frac{1}{3}\ln\frac{m_t^2}{\mu_r^2}\bigg{]}, \\
\label{gs-2} \frac{\delta g_s^{(SQCD)}}{g_s} &=&
-\frac{\alpha_s(\mu_r)}{4\pi}\bigg{[}\frac{\beta_0^{(SQCD)}}{2}\frac{1}
{\bar{\epsilon}}+\frac{N}{3}
\ln\frac{m_{\tilde{g}}^2}{\mu_r^2}+\sum_{i=1,2}\frac{1}{12}
\ln\frac{\mu^2_{\tilde{t}_i}}{\mu_r^2}\nb
\\
&+&\sum^{\tilde{q}=\tilde{u},
\tilde{d},\tilde{c},\tilde{s},\tilde{b}}_{i=1,2}
\frac{1}{12}\ln\frac{m^2_{\tilde{q}_i}}{\mu_r^2}
\bigg{]},  \nb \\
\end{eqnarray}
where the summation is taken over the squark indices of $\tilde{u}$,
$\tilde{d}$, $\tilde{c}$, $\tilde{s}$ and $\tilde{b}$. The notations
$\beta_0^{(SM-like)}$ and $\beta_0^{(SQCD)}$ are defined as
\begin{eqnarray}
\beta_0^{(SM-like)} &=& \frac{11}{3}N-\frac{2}{3}n_{lf}-\frac{2}{3},\nb \\
\beta_0^{(SQCD)} &=& -\frac{2}{3}N-\frac{1}{3}(n_{lf} +1).
\end{eqnarray}
The number of colors $N=3$, the number of light flavors $n_{lf}=5$,
and $\frac{1}{\bar{\epsilon}}=\frac{1}{\epsilon_{UV}}
-\gamma_E+\ln(4\pi)$. There are two regularization schemes,
dimensional regularization scheme and dimensional reduction
regularization scheme, customarily used in a supersymmetric gauge
theory. It is well-known the dimensional reduction regularization
scheme preserves supersymmetry at least to one-loop order,
therefore, this scheme is a proper regularization scheme for
performing the renormalization in supersymmetry. In this paper we
adopt the dimensional regularization scheme to calculate the NLO QCD
corrections. This scheme is easier than the dimensional reduction
scheme to handle in general, but violates supersymmetry because the
number of the gauge-boson and gaugino degrees of freedom are not
equal in $D \neq 4$ dimensions. To subtract the contributions of the
false, non-supersymmetric degrees of freedom and restore
supersymmetry at one-loop order, a shift between that the
$q\tilde{q}\tilde{g}$ Yukawa coupling $\widehat{g}_s$ and the
$qqg$ gauge coupling $g_s$ must be introduced \cite{shiftgs}
\begin{equation}
\label{shift} \widehat{g}_s = g_s\left [1 +
\frac{\alpha_s}{8\pi}\left (\frac{4}{3}C_A - C_F\right )\right ],
\end{equation}
where $C_A = N = 3$ and $C_F = 4/3$. Similarly the electroweak
$\tilde{t}_it\tilde{\chi}^0_1$ couplings should take a finite shift
also, which may be written as $a\widehat{e}+b\widehat{Y}_t$, with
$\widehat{e} = e[1 -{\alpha_sC_F}/{8\pi}]$ and $\widehat{Y}_t =
Y_t[1 -{3\alpha_sC_F}/{8\pi}]$ in terms of the electromagnetic
coupling $e$ and the quark-Higgs Yukawa coupling $Y_t\propto e m_t$
\cite{shiftgs}. In our numerical calculations, we take all these
shift into account and keep $\Gamma_{NLO}$ only up to ${\cal
O}(\alpha_s^2 \alpha_{ew})$.

\par
For the renormalization of the wave function and mass of top quark
by using OS scheme, we use the expressions Eqs.(10)-(13) in
Ref.\cite{top} with the replacement of $m_{\tilde{t}_i} \to
\mu_{\tilde{t}_i}$. The ${\cal O}(\alpha_s)$ QCD virtual corrections
to the decay \stttx in the MSSM can be expressed as
\begin{equation}
\Delta\Gamma_{virtual}= \frac{(2\pi)^4}{m_{\tilde{g}}} \int d
\Phi_3\overline{\sum}Re({\cal M}_{tree}{\cal
M}_{one-loop}^\dagger),
\end{equation}
where ${\cal M}_{tree}$ is the Born amplitude for the decay \stttx, and
${\cal M}_{one-loop}$ is the renormalized amplitude of all the one-loop level
Feynman diagrams involving the virtual gluon or/and gluino.

\par
\subsection{Soft gluon emission corrections}
\par
\begin{figure}[htp]
\centering
\includegraphics[scale=1.0]{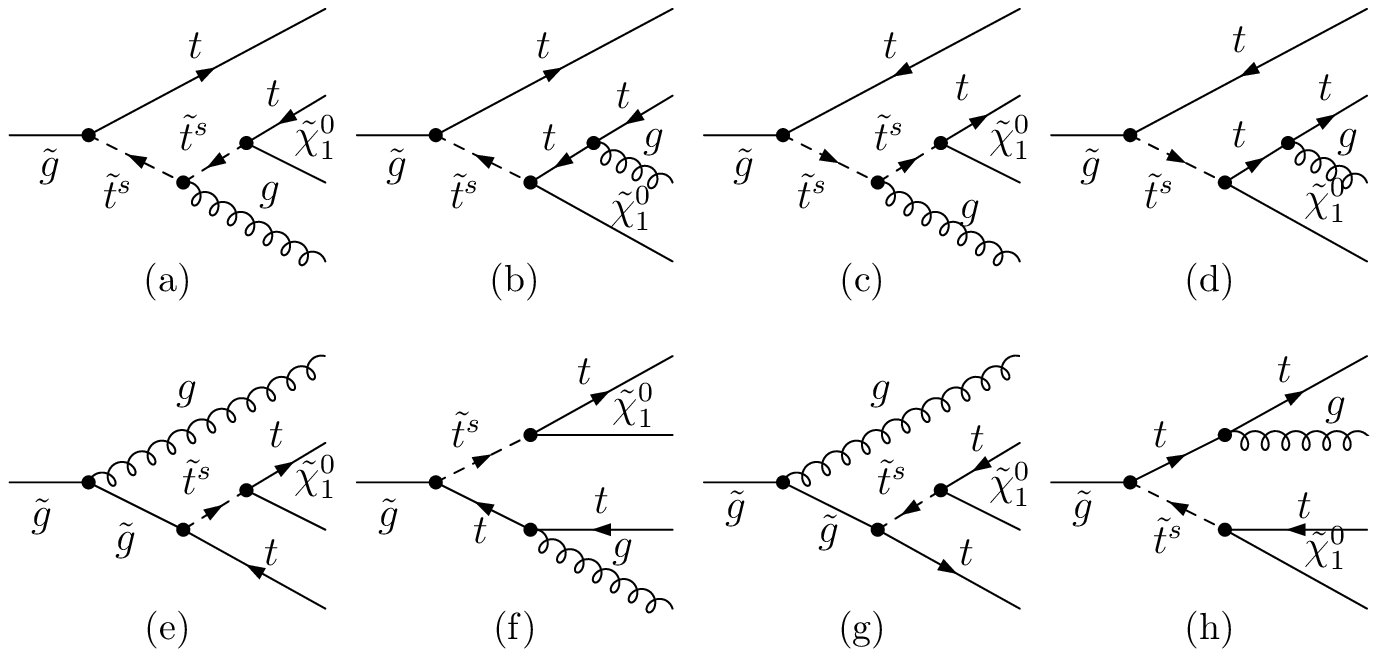}
\vspace*{0.3cm} \caption{The tree-level Feynman diagrams for the
real gluon emission decay process \stttxg. }\label{Fig3}
\end{figure}

In our calculations we denote the real gluon emission decay process
\stttxg as
\begin{equation}\label{emission}
\tilde{g}(p_1) \to t(p_2) + \bar{t}(p_3) + \tilde{\chi}^0_1(p_4) +
g(p_5),
\end{equation}
and the corresponding tree-level diagrams are shown in Fig.\ref{Fig3}.

\par
The real gluon emission decay \stttxg contains only soft IR
singularity, but no collinear IR singularity due to gluino, top
quark and top squarks being massive. This singularity can be
conveniently isolated by slicing the phase space into two different
regions defined by an arbitrary small cutoff. This method of dealing
with soft IR singularity is called phase space slicing (PSS) method
\cite{pss}. By introducing a small cutoff $\delta_s$, the phase
space of \stttxg is separated into two regions, according to whether
the emitted gluon is soft, i.e., $E_5 < \delta_s m_{\tilde{g}}/2$,
or hard, i.e., $E_5 \ge \delta_sm_{\tilde{g}}/2$. Then the decay
width of the real gluon emission process can be expressed as the
summation of the contributions over the two phase space regions,
i.e.,
\begin{eqnarray}
\Delta \Gamma_{real}(\tilde{g} \to t\bar{t}\tilde{\chi}_1^0g) =
\Delta \Gamma_{soft}(\tilde{g} \to t\bar{t}\tilde{\chi}_1^0g)+
\Delta \Gamma_{hard}(\tilde{g} \to t\bar{t}\tilde{\chi}_1^0g),
\end{eqnarray}
where $\Delta \Gamma_{soft}$ is obtained by integrating over the
soft region of the emitted gluon phase space and contains the soft
IR singularity. $\Delta \Gamma_{hard}$ is finite and can be
evaluated by using Monte Carlo technique in four dimensions
\cite{Lepage}. Although $\Delta \Gamma_{soft}$ and $\Delta
\Gamma_{hard}$ depend on the soft cutoff $\delta_s$, the total decay
width of the decay process \stttxg, $\Delta \Gamma_{real}(\tilde{g}
\to t\bar{t}\tilde{\chi}_1^0+g)$, is independent of the arbitrary
cutoff $\delta_s$. That independence is verified in our numerical
calculations. In further calculations we set $\delta_s=1\times
10^{-4}$. The decay width of the gluon bremsstrahlung \stttxg over
the soft gluon region can be expressed as \cite{ppqcd1}
\begin{eqnarray}
\Delta \Gamma_{soft} =
\Gamma_{LO}\otimes\frac{\alpha_s}{2\pi}\sum_{i,j=1\atop
i<j}^3(T_i\cdot T_j)g_{ij}(p_i,p_j),
\end{eqnarray}
where $T_i$ are the color operators
\cite{ppqcd1,Catani:1996jh,Catani:2002hc}, and $g_{ij}$ are the soft
integrals defined as \cite{ppqcd1,ppqcd2}
\begin{eqnarray}
g_{ij}(p_i,p_j) = \frac{(2\pi\mu)^{2\epsilon}}{2\pi}\int_{E_5\le
\delta_s m_{\tilde{g}}/2}\frac{d^{D-1}{\bf p_5}}{E_5}
\bigg{[}\frac{2(p_i \cdot p_j)}{(p_i \cdot p_5)(p_j \cdot
p_5)}-\frac{p_i^2} {(p_i \cdot p_5)^2}-\frac{p_j^2}{(p_j \cdot
p_5)^2}\bigg{]}.
\end{eqnarray}
By using the definitions of color operators, we get the expression
of $\Gamma_{soft}$ as
\begin{equation}
\Delta \Gamma_{soft} =
-\frac{\alpha_s}{2\pi}\bigg{[}-\frac{3}{2}(g_{12}+g_{13})
+\frac{1}{6}g_{23}\bigg{]}\Gamma_{LO}.
\end{equation}
Then the total decay width including the NLO QCD corrections of the
decay process \stttx can be obtained by summing all the contribution
parts:
\begin{equation}
\Gamma_{NLO} = \Gamma_{LO}+\Delta \Gamma_{virtual}+\Delta
\Gamma_{soft}+\Delta \Gamma_{hard}.
\end{equation}

\par
\section{Results and Discussions }
\par

In this section we present and discuss the numerical results for the
decay \stttx. we take two-loop running $\alpha_{s}$ in both LO and
NLO calculations. The number of active flavors $N_f=5$, and the QCD
parameter $\Lambda_5^{\overline{MS}}=226~MeV$. We set the
renormalization scale being equal to
$\mu=\mu_0=m_t+\frac{m_{\tilde{\chi}^0_1}}{2}$ by default. We take
the SM parameters as $\alpha_{ew}(m_Z^2) = 1/127.916$, $m_t =
172~GeV$, $m_Z = 91.1876~GeV$ and $m_W = 80.399~GeV$ \cite{jpg}.

\par
Recently, the observations at the LHC indicate that the generic
lower bound on gluino is $700~GeV$ at the $95\%$ confidence level in
simplified models containing only squarks of the first two
generations, a gluino octet and a massless neutralino
\cite{CERN-data-1}, and there are hints of the SM-like Higgs boson
with $m_h \simeq 125~GeV$ \cite{CERN-data-2}. If these hints for the
Higgs boson mass are true, then that strongly suggests the top
squarks are light. Accordingly, we consider the benchmark point
$SPS6$ scenario, which is proposed in the $SPA$ Convention and
Project \cite{SPA,snowmass,SPS}, as a numerical demonstration. The
relevant masses of SUSY particles and parameters at the $SPS6$ point
required in our numerical calculations are listed in Table
\ref{table}.
\begin{table}
\center
\begin{tabular}{ |c|c||c|c|  }
\hline
Particle & Mass(GeV) & Particle & Mass(GeV) \\
\hline \hline
$\tilde{t}_1$  & 497.89  & $\tilde{u}_R$        & 660.75 \\
$\tilde{t}_2$  & 678.64  & $\tilde{u}_L$        & 674.58 \\
$\tilde{b}_2$  & 653.05  & $\tilde{d}_R$        & 654.51 \\
$\tilde{g}$    & 721.80  & $\tilde{\chi}_1^0$   & 190.41 \\
\hline  \hline
SUSY parameter  & &SUSY parameter &  \\
\hline \hline
$M_2$          & 231.02~GeV & $\mu$                  & 391.91~GeV\\
$\tan\beta $   & 10.0       & $\theta_{\tilde{t}}$   & $58.5105^\circ$\\

\hline
\end{tabular}
\caption{Relevant SUSY parameters obtained by using ISAJET 7.82
\cite{isajet} with the input parameters at the reference point
$SPS6$. }\label{table}
\end{table}

\par
The parameters presented in Table \ref{table} are the independent
SUSY input parameters adopted in our calculations. All the other
SUSY parameters needed in this work can be determined by those in
Table \ref{table}. For example, the elements of neutralino
transformation matrix $N$ can be given by $M_2$, $\mu$, $\tan\beta$
and $m_{\tilde{\chi}_1^0}$. Since we assume $u$-, $d$-, $c$-, $s$-,
$b$-quark are all massless, the squark mixing matrices
$U^{\tilde{q}}~(q=u,d,c,s,b)$ should be all unit matrices. The
elements of top squark mixing matrix $U^{\tilde{t}}$ are determined
by $\theta_{\tilde{t}}$. In order to investigate the dependence of
the cross section on one of the parameters in Table \ref{table}, we
will vary the specific parameter while keep the other SUSY
parameters in Table \ref{table} fixed in the following calculations
except for the top squark total decay widths. From our numerical
calculations by using ISAJET 7.82 with the input parameters at the
$SPS6$ point \cite{SPA,snowmass,SPS}, we get
$\Gamma_{\tilde{t}_1}(m_{\tilde{t}_1}=497.89~GeV)=2.98~GeV$ and
$\Gamma_{\tilde{t}_2}(m_{\tilde{t}_2}=678.64~GeV)=9.70~GeV$. For
simplicity, we take the top squark decay width as
$\Gamma_{\tilde{t}_1}(m_{\tilde{t}_1})=\frac{m_{\tilde{t}_1}}{497.89~GeV}\times
2.98~GeV$
($\Gamma_{\tilde{t}_2}(m_{\tilde{t}_2})=\frac{m_{\tilde{t}_2}}{678.64~GeV}\times
9.70~GeV$) when $m_{\tilde{t}_1}$ ($m_{\tilde{t}_2}$) is variable.

\par
\begin{figure}[htb]
\centering
\includegraphics[scale=0.65]{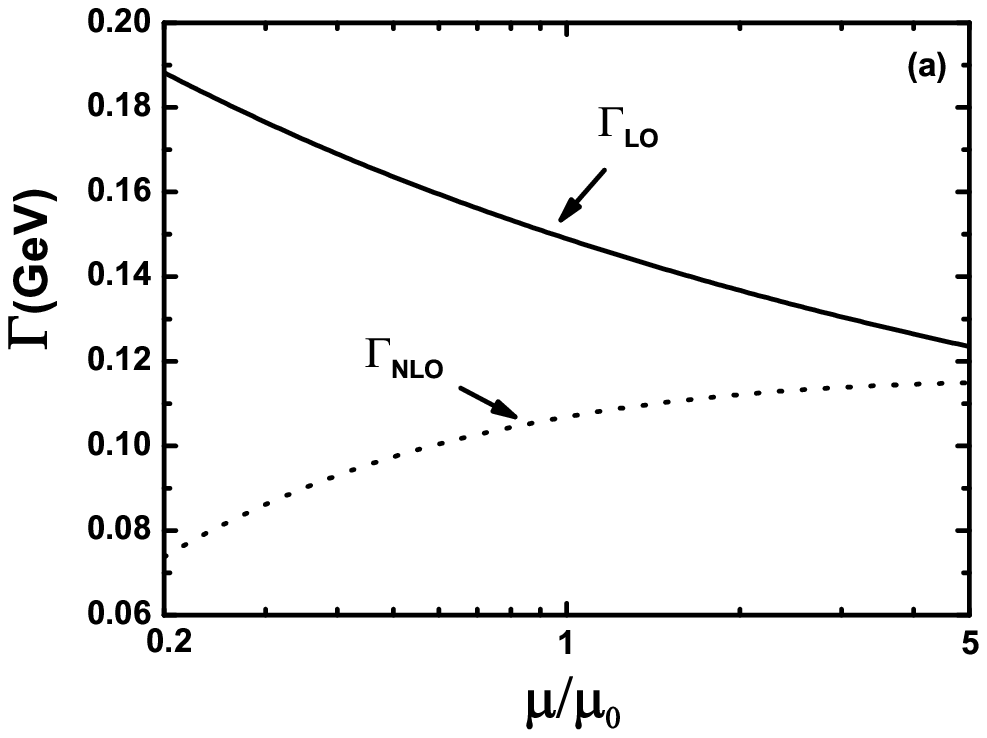}
\includegraphics[scale=0.65]{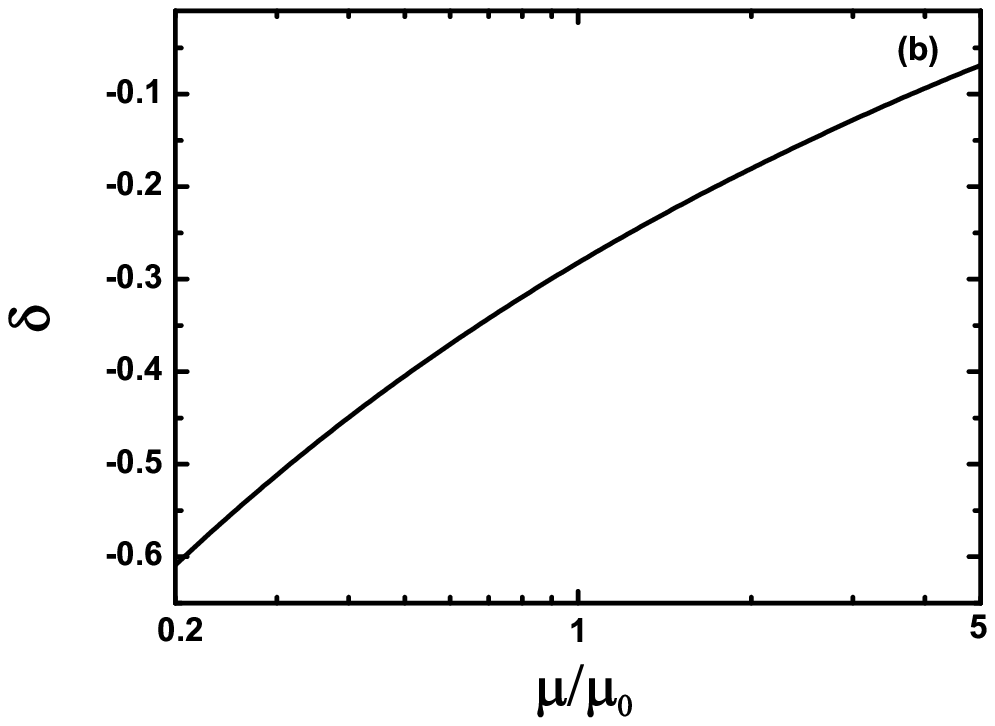}
\caption{ (a) The dependence of the LO and the QCD corrected \stttx
decay width on the renormalization scale. (b) The dependence of the
relative correction for the \stttx decay width on the
renormalization scale.  } \label{Fig4-1}
\end{figure}

We show the LO, NLO QCD corrected \stttx decay widths and the
corresponding relative correction ($\delta \equiv
\frac{\Gamma_{NLO}-\Gamma_{LO}}{\Gamma_{LO}}$) as the functions of
the renormalization scale in Figs.\ref{Fig4-1}(a) and (b)
separately, where we denote $\mu = \mu_r$ and $\mu_0=
m_t+m_{\tilde{\chi}_1^0}/2$. If we define the scale uncertainty for
the \stttx process as $\eta = \frac{|\Gamma
(\mu=5\mu_0)-\Gamma(\mu=0.2\mu_0)|}{\Gamma(\mu=\mu_0)}$, from the
curves in Fig.\ref{Fig4-1}(a) we can figure out the corresponding
uncertainties being $\eta_{LO}=43.0\%$ and $\eta_{NLO}=38.4\%$. It
is clear that the LO decay width is strongly related to the
renormalization scale in the plotted $\mu$ range, while the NLO QCD
corrections obviously reduce the scale uncertainties.
Fig.\ref{Fig4-1}(b) shows that the relative QCD correction $\delta$
raises from $-60.7\%$ to $-6.9\%$ when the renormalization scale
goes from $0.2\mu_0$ to $5\mu_0$, and the relative correction has
the value of $-28.2\%$ at the location of $\mu=\mu_0$.

\par
\begin{figure}[htb]
\centering
\includegraphics[scale=0.7]{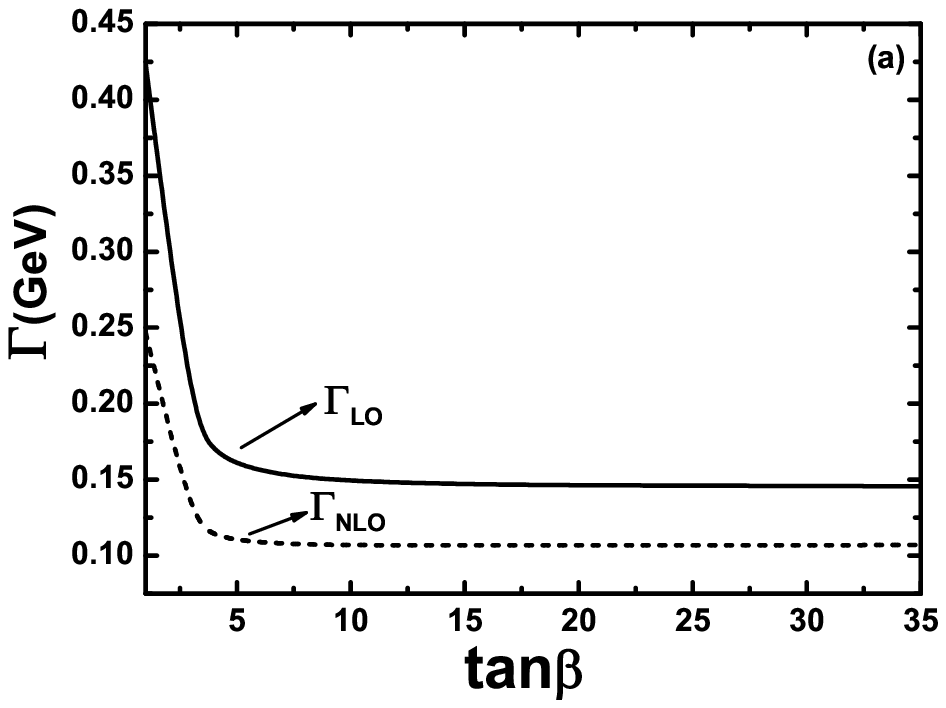}
\includegraphics[scale=0.7]{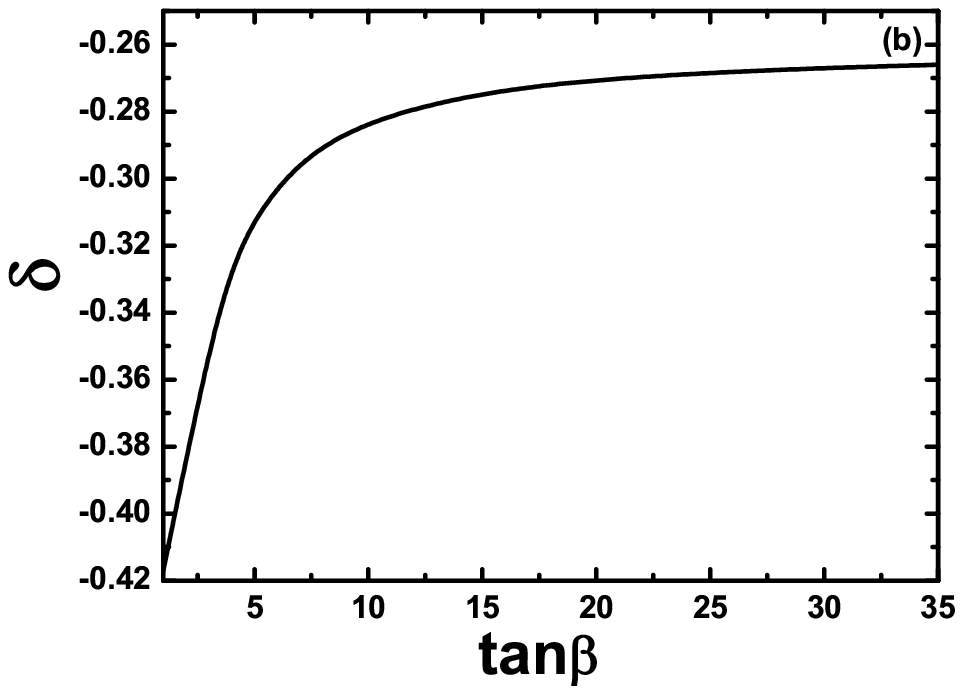}
\caption{ (a) The LO and the QCD NLO corrected decay widths of
\stttx as the functions of $\tan\beta$. (b) The corresponding
relative correction as the function of $\tan\beta$. } \label{Fig4}
\end{figure}

The decay widths of \stttx at ${\cal O}(\alpha_s \alpha_{ew})$ and
up to ${\cal O}(\alpha_s^2 \alpha_{ew})$, and the corresponding
relative correction($\delta \equiv
\frac{\Gamma_{NLO}-\Gamma_{LO}}{\Gamma_{LO}}$) as functions of the
ratio of the vacuum expectation values (VEVs) $\tan\beta$ are
depicted in Fig.\ref{Fig4}(a) and Fig.\ref{Fig4}(b) separately, with
$\tan\beta$ varying from 1 to 35 and the other SUSY input parameters
being adopted from the reference point ${\rm SPS6}$ in Table
\ref{table}. The curves for the decay widths $\Gamma_{LO}$ and
$\Gamma_{NLO}$ drawn in Fig.\ref{Fig4}(a) demonstrate that the NLO
QCD correction suppresses the LO decay width of the decay process
\stttx. Both the decay widths $\Gamma_{LO}$ and $\Gamma_{NLO}$
quantitatively fall down when $\tan\beta$ goes up from 1 to 5, and
vary smoothly when $\tan\beta > 5$. Fig.\ref{Fig4}(b) shows the
relative correction raises from  $-41.7\%$ to $-26.6\%$ with the
increment of $\tan\beta$ from $1$ to $35$.

\vskip 5mm
\begin{figure}[htb]
\centering
\includegraphics[scale=0.7]{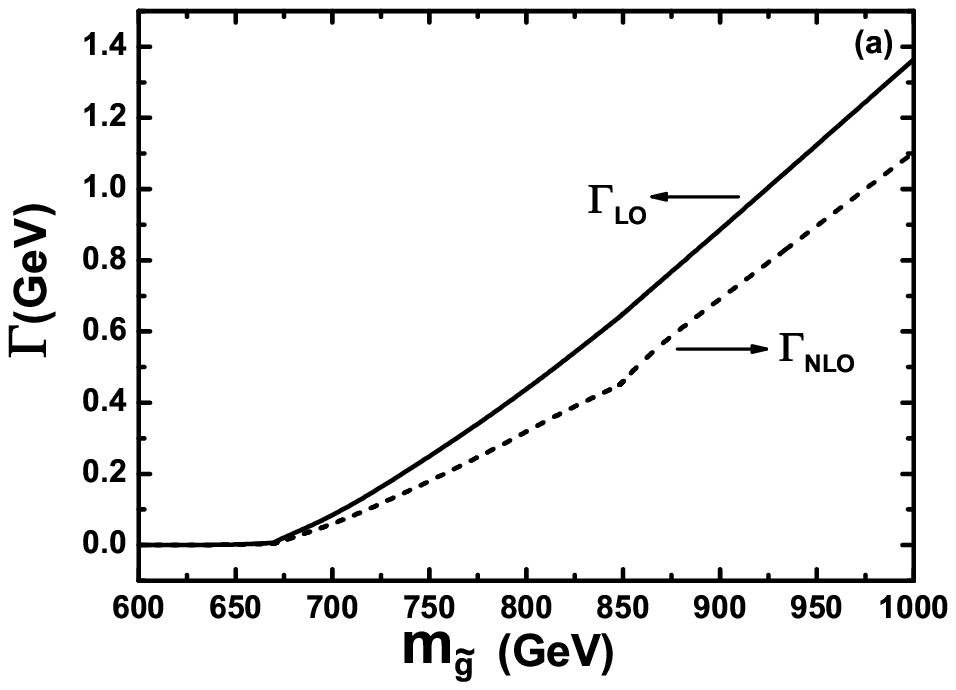}
\includegraphics[scale=0.7]{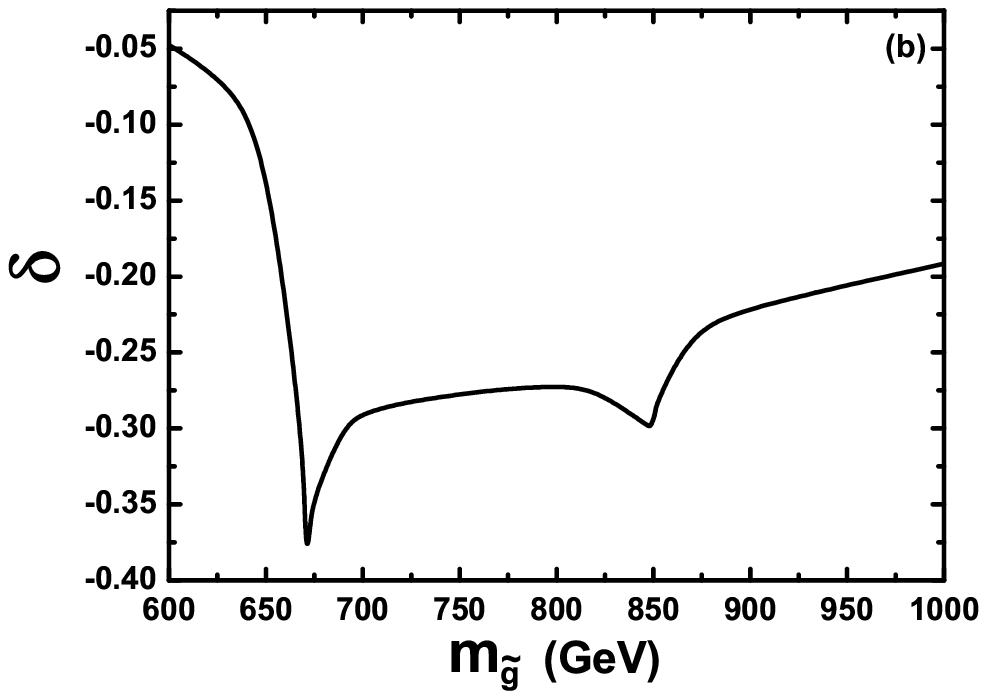}
\caption{ (a) The LO and the QCD NLO corrected decay widths of
\stttx as the functions of gluino mass $m_{\tilde{g}}$. (b) The
corresponding relative correction as the function of
$m_{\tilde{g}}$.} \label{Fig5}
\end{figure}

\vskip 5mm
\begin{figure}[htb]
\centering
\includegraphics[scale=0.7]{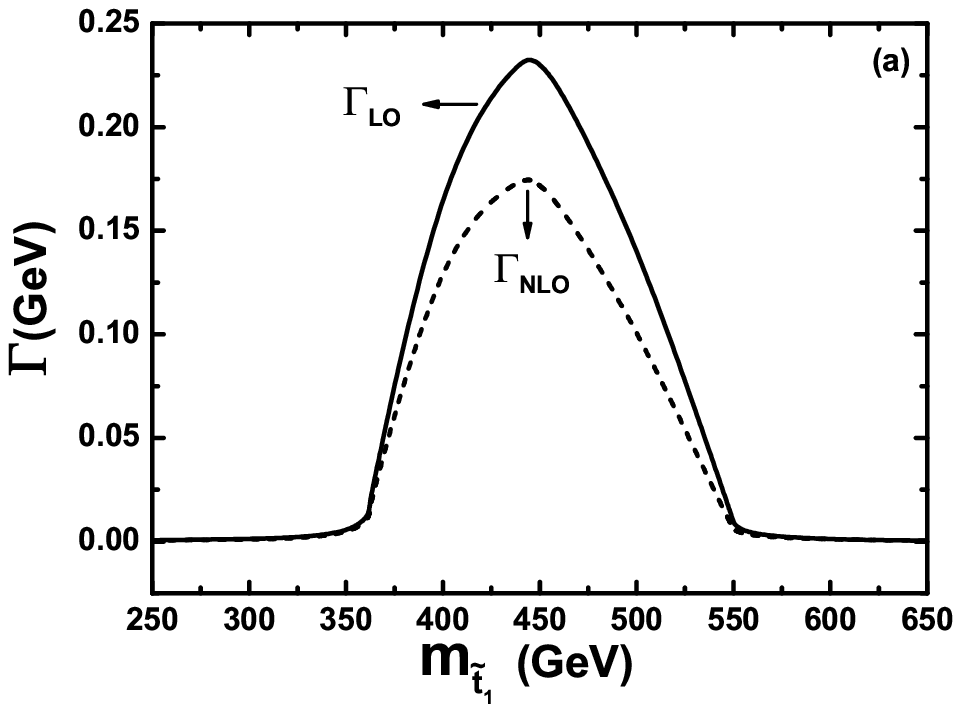}
\includegraphics[scale=0.7]{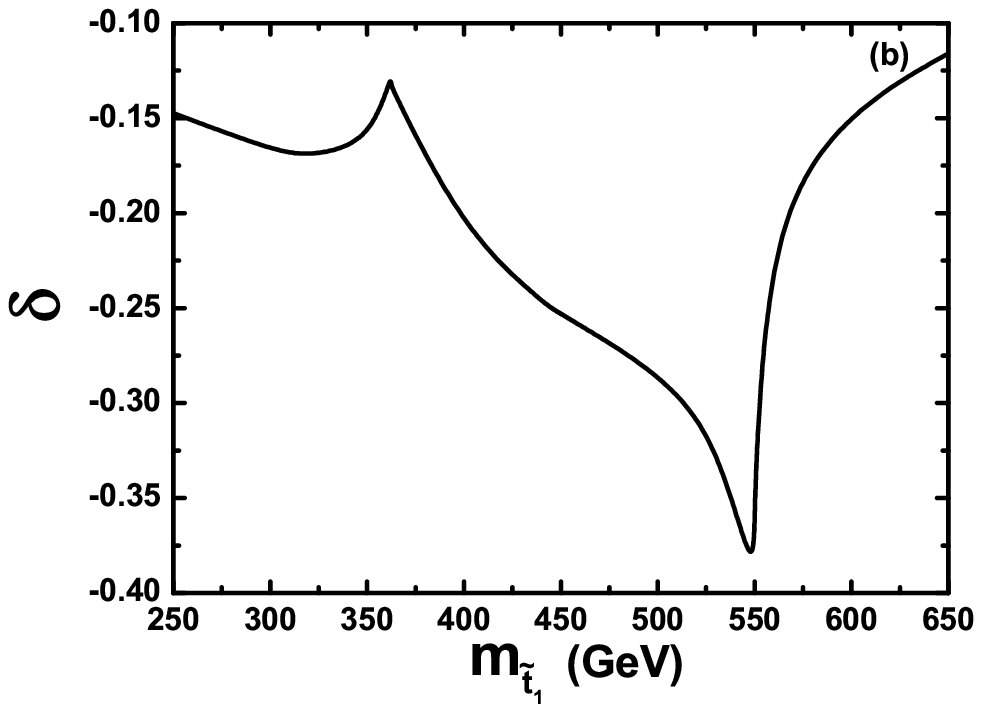}
\centering \caption{ (a) The LO and the QCD NLO corrected decay widths
as the functions of $m_{\tilde{t}_1}$. (b) The corresponding relative
correction as the function of $m_{\tilde{t}_1}$.} \label{Fig6}
\end{figure}

\par
The dependences of the LO, NLO QCD corrected decay widths of \stttx
and the corresponding relative correction on the gluino mass
$m_{\tilde{g}}$ are depicted in Figs.\ref{Fig5}(a-b), with the
parameter $m_{\tilde{g}}$ in the range of $[600,~1000]~GeV$ and the
other SUSY input parameters being adopted from the reference point
${\rm SPS6}$ in Table \ref{table}. Fig.\ref{Fig5}(a) shows that the
NLO QCD corrections are negative, and the LO and NLO QCD corrected
decay widths at the position of $m_{\tilde{g}}=721.8~GeV~(1000~GeV)$
can reach $0.1490~GeV~(1.364~GeV)$ and $0.1069~GeV~(1.103~GeV)$,
respectively. In Fig.\ref{Fig5}(a) there exist two knees on the NLO
curve at the positions satisfying the conditions of
$m_{\tilde{g}}=m_{\tilde{t}_1}+m_t=669.9~GeV$ and
$m_{\tilde{g}}=m_{\tilde{t}_2}+m_t=850.6~GeV$, but this resonance
effect is not observable on the LO curve. Fig.\ref{Fig5}(b) shows
that the relative corrections at the points of
$m_{\tilde{g}}=721.8~GeV$ and $m_{\tilde{g}}=1000~GeV$ are $-28.2\%$
and $-19.2\%$, respectively. We can see obviously from
Fig.\ref{Fig5}(b) that the curve for the relative correction has two
obvious spikes at the vicinities of $m_{\tilde{g}}=699.9~GeV$ and
$m_{\tilde{g}}=850.6~GeV$ due to the top squark resonance effects.

\par
In Fig.\ref{Fig6}(a) we plot the LO and the NLO QCD corrected decay
widths, $\Gamma_{LO}$ and $\Gamma_{NLO}$, as the functions of
$\tilde{t}_1$ mass by keeping the other SUSY input parameters at the
reference point ${\rm SPS6}$ shown in Table \ref{table} except the
$\Gamma_{\tilde{t}_1}$ varying with $m_{\tilde{t}_1}$. The
corresponding relative NLO QCD correction ($\delta$) as the function
of $m_{\tilde{t}_1}$ is shown in Fig.\ref{Fig6}(b).
Fig.\ref{Fig6}(a) shows that the curves for $\Gamma_{LO}$ and
$\Gamma_{NLO}$ are significantly enhanced in the $m_{\tilde{t}_1}$
range of $[362.4,~ 549.8]~GeV$ due to the resonance effect of the
intermediate on-shell $\tilde{t}_1$ in the decay chains of
$\tilde{g} \to \bar{t} \tilde{t}_1 \to \bar{t} t \tilde{\chi}_1^0$
and $\tilde{g} \to t \bar{\tilde{t}_1} \to t \bar{t}
\tilde{\chi}_1^0$, where the two turning points satisfy the
conditions of $m_{\tilde{t}_1}=m_{\tilde{\chi}_1^0}+m_t=362.4~GeV$
and $m_{\tilde{t}_1}=m_{\tilde{g}}-m_t=549.8~GeV$. The LO and the
NLO QCD corrected decay widths reach their maxima of $0.2341~GeV$
and $0.1760~GeV$ at the position of $m_{\tilde{t}_1}=440~GeV$,
respectively. On the curve of the relative correction in
Fig.\ref{Fig6}(b) there exist two spikes at the vicinities of
$m_{\tilde{t}_1}=362.4~GeV$ and $m_{\tilde{t}_1}=549.8~GeV$, which
also reflects the resonance of the intermediate on-shell
$\tilde{t}_1$ in the decay process $\tilde{g} \to t \bar{t}
\tilde{\chi}_1^0$.

\par
\begin{figure}[htb]
\centering
\includegraphics[scale=0.7]{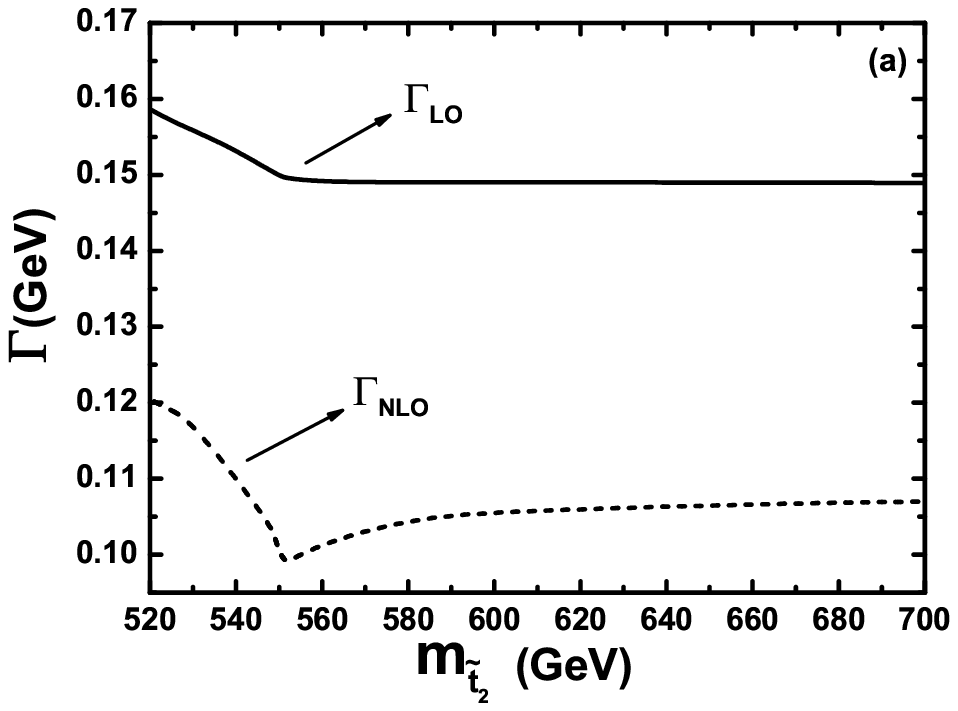}
\includegraphics[scale=0.7]{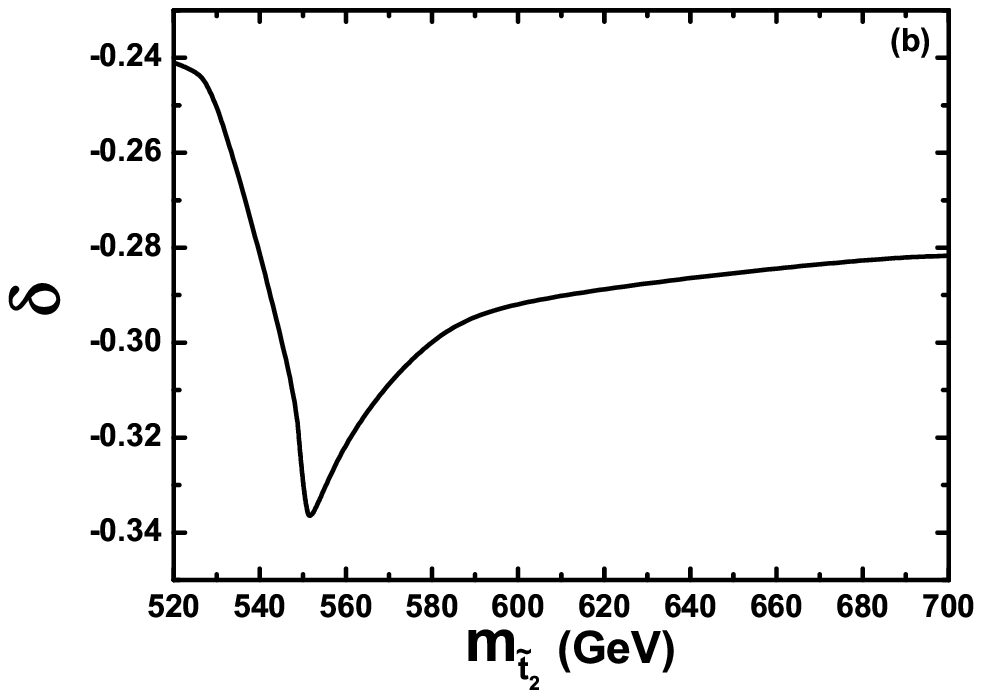}
\centering \caption{ (a) The LO and the NLO QCD corrected decay widths
as the functions of $m_{\tilde{t}_2}$. (b) The corresponding relative
correction as the function of $m_{\tilde{t}_2}$. } \label{Fig7}
\end{figure}

The LO, NLO QCD corrected decay widths and the corresponding
relative NLO QCD correction as the functions of $m_{\tilde{t}_2}$
are depicted in Fig.\ref{Fig7}(a) and Fig.\ref{Fig7}(b),
respectively, with $m_{\tilde{t}_2}$ varying in the range of $[520,
~700]~GeV$. In Figs.\ref{Fig7}(a,b) all the SUSY input parameters
are taken from the reference point ${\rm SPS6}$ except for
$m_{\tilde{t}_2}$ and $\Gamma_{\tilde{t}_2}$. Fig.\ref{Fig7}(a)
shows that the LO and the NLO QCD corrected decay widths have the
values of $0.1586~GeV~(0.1490~GeV)$ and $0.1204~GeV~(0.1070~GeV)$ at
$m_{\tilde{t}_2}=520~GeV~(700~GeV)$, respectively. We can see from
this figure that in the region of $m_{\tilde{t}_2}\leq 549.8~ GeV$
both $\Gamma_{LO}$ and $\Gamma_{NLO}$ have relative large values due
to the resonance effect of the intermediate on-shell $\tilde{t}_2$
(where the condition $m_{\tilde{t}_2}\leq
m_{\tilde{g}}-m_t=549.8~GeV$ is satisfied). From Fig.\ref{Fig7}(b)
we can read out that the relative corrections are $-24.1\%$ and
$-28.2\%$ at the positions of $m_{\tilde{t}_2}=520~GeV$ and
$m_{\tilde{t}_2}=700~GeV$, respectively. Again, there is a spike on
the curve for relative correction at the vicinity of
$m_{\tilde{t}_2} = 549.8~GeV$ induced by the resonance effect of the
intermediate on-shell $\tilde{t}_2$. \vskip 5mm

\par
\begin{figure}[htb]
\centering
\includegraphics[scale=0.7]{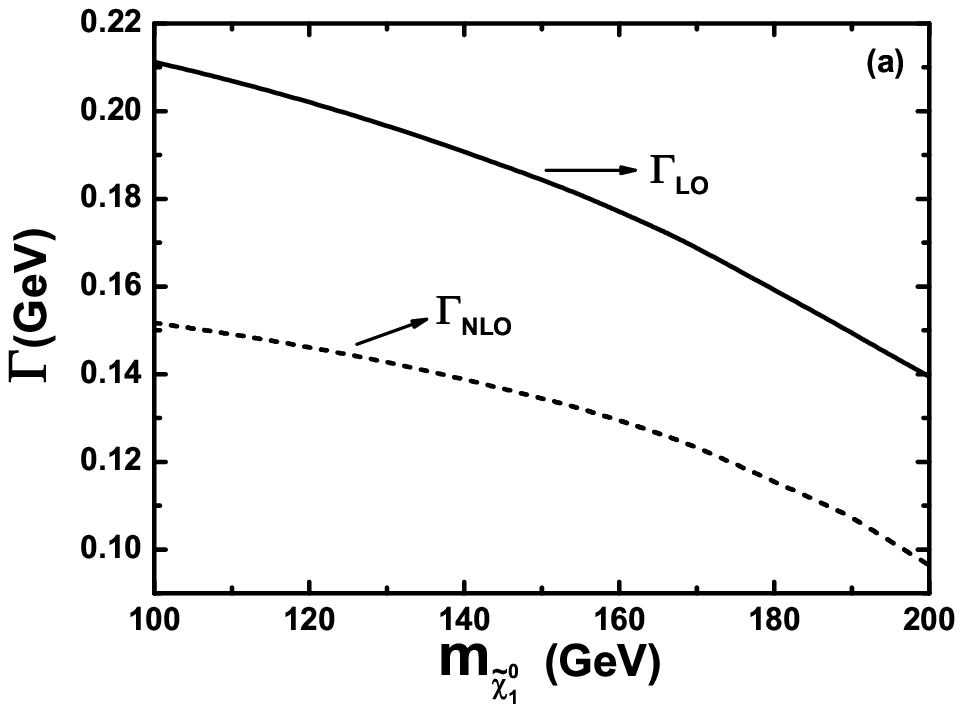}
\includegraphics[scale=0.7]{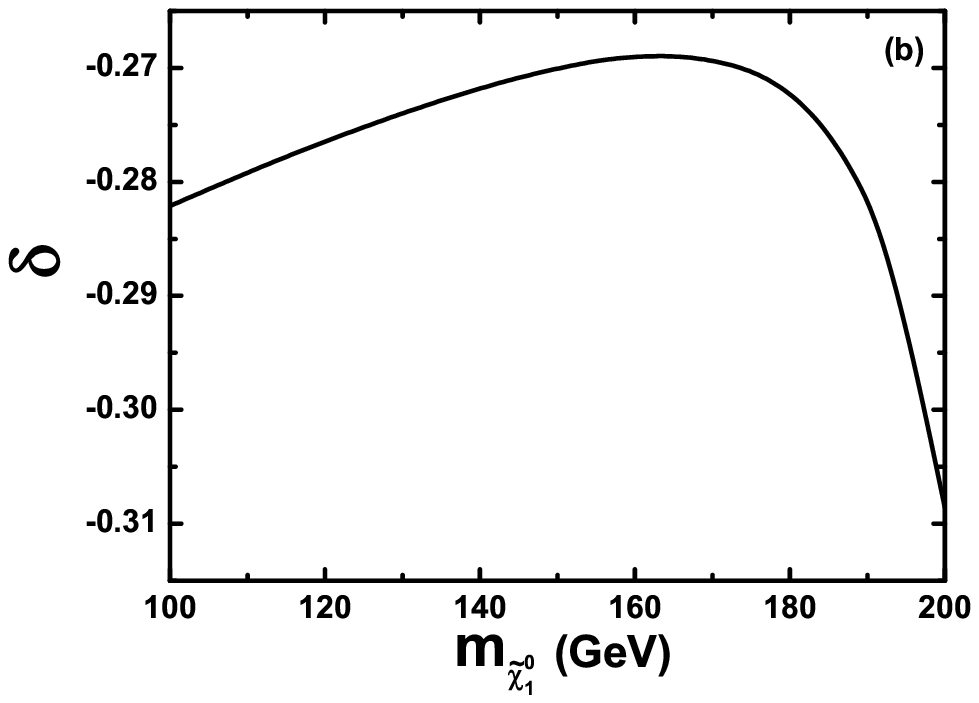}
\caption{ (a) The LO and the QCD NLO corrected decay
widths of \stttx as the functions of $m_{\tilde{\chi}_1^0}$. (b) The corresponding
relative correction as the function of $m_{\tilde{\chi}_1^0}$. } \label{Fig8}
\end{figure}

The decay widths of \stttx at the LO and the QCD NLO, and the
corresponding relative correction versus $m_{\tilde{\chi}_1^0}$ are
shown in Fig.\ref{Fig8}(a) and Fig.\ref{Fig8}(b) separately, with
$m_{\tilde{\chi}_1^0}$ being in the range from $100~GeV$ to
$200~GeV$ and the other SUSY parameters being taken from the
reference point ${\rm SPS6}$ shown in Table \ref{table}. The curves
for the LO and the NLO QCD corrected decay widths decrease with the
increment of $m_{\tilde{\chi}_1^0}$. The curve for $\Gamma_{LO}$
$(\Gamma_{NLO})$ goes down from $0.211~GeV~(0.152~GeV)$ to
$0.139~GeV~(0.0964~GeV)$ when $m_{\tilde{\chi}_1^0}$ varies from
$100~GeV$ to $200~GeV$. The corresponding relative corrections at
the points of $m_{\tilde{\chi}_1^0}=100~GeV$ and
$m_{\tilde{\chi}_1^0}=200~GeV$ are $-28.2\%$ and $-30.9\%$,
respectively.

\par
\begin{figure}[htb]
\centering
\includegraphics[scale=0.7]{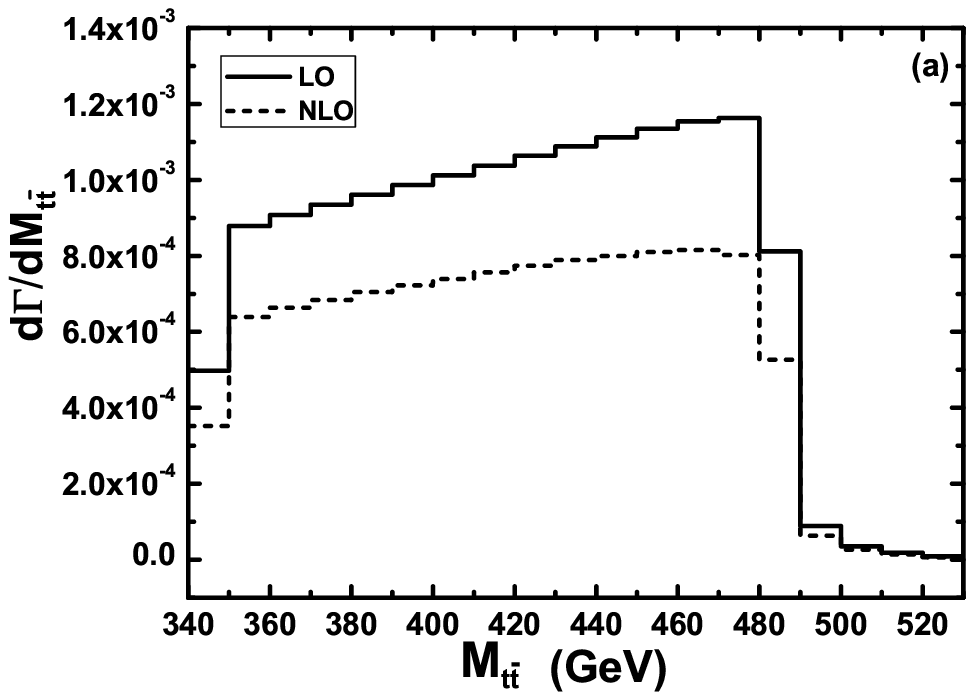}
\includegraphics[scale=0.7]{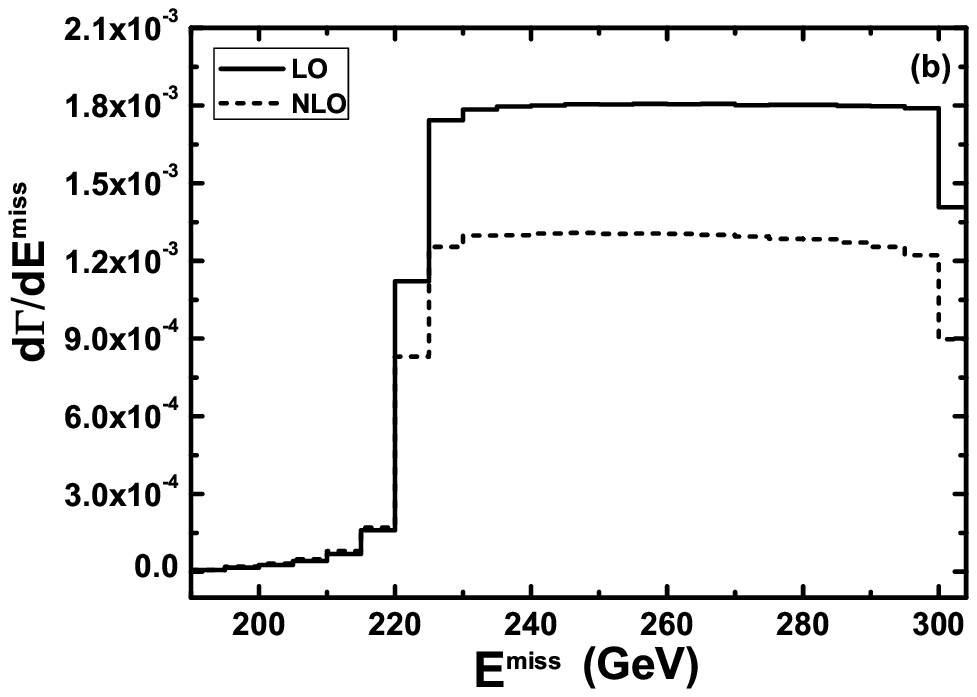}
\caption{ (a) The $t\bar t$ invariant mass distributions at the LO
and the QCD NLO ($\frac{d\Gamma_{LO}}{dM_{t\bar t}}$,
$\frac{d\Gamma_{NLO}}{dM_{t\bar t}}$) for the ${\rm SPS6}$ parameter
set in Table \ref{table}. (b) The missing energy distributions at
the LO and the QCD NLO ($\frac{d\Gamma_{LO}}{dE^{miss}}$,
$\frac{d\Gamma_{NLO}}{dE^{miss}}$) with the ${\rm SPS6}$ parameter
set listed in Table \ref{table}. } \label{Fig9}
\end{figure}

The LO and the QCD NLO corrected distributions of the top-pair
($t\bar t$) invariant mass $M_{t\bar t}$ and the missing energy
$E^{miss}$ ($\frac{d\Gamma_{LO}}{dM_{t\bar t}}$,
$\frac{d\Gamma_{NLO}}{dM_{t\bar t}}$,
$\frac{d\Gamma_{LO}}{dE^{miss}}$ and
$\frac{d\Gamma_{NLO}}{dE^{miss}}$) at the reference point ${\rm
SPS6}$ are shown in Fig.\ref{Fig9}(a) and Fig.\ref{Fig9}(b),
respectively. The kinematics of the decay process \stttx at this
point constraints the $\frac{d\Gamma_{LO,NLO}}{dM_{t\bar t}}$
distributions in the range between $M_{t\bar t}^{min}=2m_t\simeq
340~GeV$ and $M_{t\bar
t}^{max}=m_{\tilde{g}}-m_{\tilde{\chi}_1^0}\simeq 530~GeV$ as shown
in Fig.\ref{Fig9}(a), and the $\frac{d \Gamma_{LO,NLO}}{d E^{miss}}$
distributions are limited in the range between
$E^{miss}_{min}=m_{\tilde{\chi}_1^0}\simeq 190~GeV$ and
$E^{miss}_{max}= \frac{m_{\tilde{g}}}{2} - \frac{(4m_t^2 -
m_{\tilde{\chi}_1^0}^2)}{2m_{\tilde{g}}}\simeq 304~GeV$ as shown  in
Fig.\ref{Fig9}(b). We compare these differential decay widths and
find that the NLO QCD corrections suppress the corresponding LO
differential decay widths $\frac{d\Gamma_{LO}}{dM_{t\bar t}}$ and
$\frac{d\Gamma_{LO}}{dE^{miss}}$ significantly, but do not change
obviously the line shapes of the LO differential decay widths. From
Fig.\ref{Fig9}(a) we see that the $M_{t\bar t}$ distributes mainly
in the range of $[350,~ 480]~ GeV$, and the differential decay
widths raise slowly with the increment of $M_{t\bar t}$ in this
region. Fig.\ref{Fig9}(b) shows that the missing energy events are
mainly concentrated and nearly uniformly distributed in the
$E^{miss}$ range of $[225,~ 300]~ GeV$ for both the LO and QCD NLO
corrected distributions. \vskip 5mm

\par
\section{Summary}
\par
In this paper we calculate the NLO QCD corrections to the decay
process \stttx in the MSSM. As a numerical demonstration we present
and discuss the NLO QCD corrections around the ${\rm SPS6}$
benchmark point. We find that the NLO QCD corrections significantly
suppress the corresponding LO decay width, and at the scenario ${\rm
SPS6}$ point the LO and the NLO QCD corrected decay widths have the
values of $0.1490~GeV$ and $0.1069~GeV$ respectively, and the
corresponding relative correction is $-28.2\%$. We analyze the
dependence of the NLO QCD corrected decay width and the
corresponding relative correction on the ratio of the vacuum
expectation values, gluino mass, scalar top-quark masses and the
lightest neutralino mass, respectively, around the scenario point
${\rm SPS6}$. Our numerical results show that the absolute NLO QCD
relative correction can exceed $30\%$ in our chosen parameter space.
Therefore, it is necessary to take the NLO QCD corrections into
account for the precise experimental measurement at future
colliders. We compare the distributions of the $t\bar t$ invariant
mass and the missing energy at the LO and the QCD NLO and find that
the line shapes of the differential decay widths at the LO,
$\frac{d\Gamma_{LO}}{dM_{t\bar t}}$ and
$\frac{d\Gamma_{LO}}{dE^{miss}}$, are not obviously changed by the
NLO QCD corrections.

\vskip 5mm
\par
\noindent{\large\bf Acknowledgments:} This work was supported in
part by the National Natural Science Foundation of China (Contract
No.11075150, No.11005101), and the Specialized Research
Fund for the Doctoral Program of Higher Education (Contract
No.20093402110030).

\end{document}